\documentclass[english,keywords,aps,twocolumn]{revtex4-1}
\usepackage{babel}
\usepackage{amssymb}
\usepackage{graphicx}
\usepackage{color}
\usepackage{bm}
\usepackage{longtable}
\usepackage{amsmath} 
\usepackage{amsfonts}
\usepackage{amssymb}
\usepackage{amsmath}%
\usepackage{amsfonts}%
\usepackage{amssymb}%
\usepackage[cal=cm]{mathalfa}
\usepackage{siunitx}
\usepackage{graphicx}
\usepackage{braket}
\usepackage{sidecap}
\usepackage{color}
\usepackage{bbm}
\usepackage{nicefrac}
\usepackage{float}
\usepackage[dvipsnames]{xcolor}
\usepackage[utf8]{inputenc}
\usepackage{xcolor, soul} 
\usepackage{sidecap}

\usepackage{physics}
\usepackage{amsmath,amssymb,amsthm,mathrsfs,amsfonts,dsfont} 
\usepackage{braket}

\usepackage{bbold}

\usepackage{fancyhdr} 

\usepackage{hyperref} 

\newcommand{\angstrom}{\textup{\AA}}

\begin{document}

\title{Direct experimental test of commutation relation\\ via weak value}

\author{Richard Wagner$^{1}$}
\author{Wenzel Kersten$^{1}$}
\author{Armin Danner$^{1}$}
\author{Hartmut Lemmel$^{1,2}$}
\author{Alok Kumar Pan$^{3}$}
\author{Stephan Sponar$^{1}$}
\email{stephan.sponar@tuwien.ac.at}

\affiliation{%
$^1$Atominstitut, TU Wien, Stadionallee 2, 1020 Vienna, Austria \\
$^2$Institut Laue-Langevin, 38000, Grenoble, France\\
$^3$National Institute of Technology Patna, Ashok Rajhpath, Patna 800005, India}

\begin{abstract}

The canonical commutation relation is the hallmark of quantum theory and Heisenberg's uncertainty relation is a direct consequence of it. But despite its fundamental role in quantum theory, surprisingly, its genuine direct experimental test has hitherto not been performed. In this article, we present a novel scheme to directly test the canonical commutation relation between two dichotomic observables, by exploiting the notion of weak measurement. The imaginary part of a suitably formulated weak value enables this direct test. The measurement of the weak value of a path-qubit observable in a neutron interferometer experiment is used to verify the approach. The experiment is realized using a newly developed technique in our neutron interferometric setup where the neutron's spin/energy degree of freedom serves as ancilla.

\end{abstract}

\maketitle

\section{Introduction}
In his epoch-making paper ``Quantum mechanical reinterpretation of kinematic and mechanical relations'' in July 1925, Heisenberg's \cite{heisenberg25}  put forward his breakthrough idea by introducing an entirely new representation of the position variable in terms of a set of transition amplitudes corresponding to atomic radiation. This led him to propose an unfamiliar rule of multiplication of two amplitudes in order to obtain correct intensities. It was immediately identified as matrix multiplication by Born. In the same year, by introducing a mathematically elegant language of matrices, Born and Jordan \cite{born25} systematically formalized the Heisenberg's `matrix mechanics' and provided the world a one-line epitaph
\begin{align}
\label{com}
	\hat{p}\hat{q}-\hat{q} \hat{p}=\rm{i}\hbar\, \mathbb{1}, 
\end{align}
where $\hat{p}$ and $\hat{q}$ are the matrix forms of position and momentum variables in classical mechanics. Equation (\ref{com}) is now widely known as canonical  commutation relation.
The very next year, Schr$\ddot{\rm o}$dinger \cite{sch} surprisingly put forward an alternative theory, coined as `wave mechanics', in that the core element is the wave function $\Psi$. In terms of interpretation and spirit, the wave mechanics greatly differs from matrix mechanics but produces equivalent quantum statistics.

Two years after his first breakthrough work, in another seminal paper \cite{heisenberg27}, Heisenberg proposed his famous uncertainty relation $\delta p \delta q \sim h$ which he regarded as a direct mathematical consequence of the canonical commutation rule in Eq.\,(\ref{com}). Here $\delta p$ and $\delta q$ are sort of uncertainties in momentum and position measurements respectively. Later, building upon Kennard's \cite{Kennard27} idea of interpreting the uncertainties as standard deviation,  Robertson \cite{rob29} generalized Heisenberg's preparation uncertainty relation for any two arbitrary observables $A$ and $B$ so that $\Delta A\Delta B\geq |\langle [A,B]\rangle|/2$. In recent times, the distinction between preparation and measurement uncertainty relations has been made and many interesting new formulations have also been proposed \cite{arthurs65,busch85, ozawa03, busch13, busch14, buscemi14,cyril13}. Quite a few of them have experimentally been tested \cite{erhart12,steinberg12,sponar15,sulyok,demirel19}. However, no clear consensus among physicists as to the appropriate measure of measurement (in)accuracies has been reached till date \cite{ozawa03,busch13}. 

 Although the uncertainty relations are regarded as direct (or indirect) consequences of relevant commutation relations, only very few attempts have been made to directly test the commutation relations \cite{Hasegawa1997, Wagh1997, Kim2010}.  We attribute this to the fact that the joint probability of two non-commuting observables does not exist in quantum theory and the product of such observables is in general not Hermitian. Hence, a direct test of commutation relation is non-trivial in experiment. In this work, we propose a theoretical scheme in which a single anomalous weak value enables a direct test of the commutation relation between qubit observables. Further, we report an experimental test using neutron interferometry.   

The weak value of an observable arises in a novel conditional measurement protocol  \cite{aav}, widely known as `weak measurement'.  
Consider a system prepared in a state $|\psi_{\mathrm{i}}\rangle$ (commonly known as pre-selected state) and an observable $\hat{A}$ to be measured on the system. If the measurement interaction between the system and the apparatus is weak, the system state remains grossly undisturbed. If a particular outcome $|\psi_{\mathrm{f}}\rangle$ is selected after such a weak interaction by sequentially performing a strong measurement (post-selection), the final pointer state yields the weak value, quantified by the formula \cite{aav} 
\begin{equation}
\langle A\rangle_{\mathrm{w}}^{\psi_{\mathrm{i}}, \psi_{\mathrm{f}}}=\frac{\left\langle \psi_{\mathrm{f}}|A|\psi
_{\mathrm{i}}\right\rangle }{\left\langle \psi_{\mathrm{f}}\right\vert \psi_{\mathrm{i}}\rangle
}\label{aavwvalue}.
\end{equation}
Unlike the expectation value $\langle A\rangle $, the weak value $\langle A\rangle_{\mathrm{w}}^{\psi_{\mathrm{i}}, \psi_{\mathrm{f}}}$ can be beyond the range of eigenvalues and can even be complex. The physical interpretation and implications of complex and large weak values have been widely discussed in the literature (see, for example, \cite{dressel}).  

A flurry of theoretical \cite{duck,av91,resch,pusey,dressel,vaid13,ah13,pigeon1} and experimental works \cite{ritchie,pryde,hosten,jwvexp,starling,steinberg11,lundeen11,danan,piacentini,denk, xu,david} on weak measurement have been reported in the last two decades. An anomalous weak value is proven to be beneficial in many practical applications, such as identifying the tiny spin Hall effect \cite{hosten}, detecting very small beam deflections \cite{starling} and improving precision in metrology \cite{xu,david}. Besides, it provided new insights into conceptual foundations of quantum theory \cite{av91,ah13,vaid13, pusey,pigeon1} and experiments have been performed to demonstrate for observing quantum trajectories for photons \cite{steinberg11},  and realizing counter-intuitive quantum paradoxes  \cite{jwvexp,denk, pigeon2}. 

In a fundamental experiment by Lundeen \emph{et al.} \cite{lundeen11}, a direct measurement of the quantum wave function $\Psi(x)$ was performed by using weak measurements. Post-selecting the system in momentum state $|p\rangle$, the measured weak value of position $\Pi_{x}=|x\rangle\langle x|$ becomes proportional to $\Psi(x)$. Along the same vein, we formulate a novel scheme in this article so that a single (anomalous) weak value enables a direct experimental test of the canonical commutation relation between two observables in a qubit system. We experimentally performed the measurement of the weak value of a path-qubit observable in our neutron interferometer setup where the spin/energy degree of freedom serves as a pointer. The experimental results are in good agreement with our theoretical prediction and hence provide a genuine experimental verification of the canonical commutation relation.

\section*{Theory}
Without loss of generality, consider two  non-commuting qubit observables, say $\hat{A}$ and $\hat{B}$ satisfying $\langle \psi|[\hat{A}, \hat{B}]|\psi\rangle \neq \gamma$ for all $|\psi\rangle$ where $\gamma$ is a nonzero quanity.  Since the product  $\hat{A}\hat{B}$ may not be Hermitian in general, the traditional von Neumann quantum measurement scheme cannot be carried but the weak measurement suffices for our purpose. Non-Hermitian observables may be measured \cite{nirala2019} using weak measurements. But our scheme here is direct and fundamentally different, as follows. If $\Pi_{A}^{+} =|+_{A}\rangle\langle +_{A}|$ and $\Pi_{B}^{+}=|+_{B}\rangle\langle +{_{B}}|$ are positive-eigenvalue projectors of $\hat{A}$ and $\hat{B}$, respectively, then by writing $\hat{A}=2\Pi_{A}^{+}-\mathbb{1}$ and $\hat{B}=2\Pi_{B}^{+}-\mathbb{1}$, one has 
\begin{eqnarray}
\label{ab0}
	\langle \psi|[\hat{A}, \hat{B}]|\psi\rangle&=&4 \langle \psi| \Pi_{A}^{+}\Pi_{B}^{+}|\psi\rangle - 4\langle \psi|\Pi_{B}^{+}\Pi_{A}^{+} |\psi\rangle \\
\nonumber
	&=&-4 |\langle +_{B}|\psi\rangle|^{2}\Big(\langle \Pi_{A}^{+}\rangle^{\psi,+_{B}}_{\mathrm{w}} - \langle \Pi_{A}^{+}\rangle^{\psi,+_{B}\,\ast}_{\mathrm{w}}\Big)\\	
\nonumber
&=& -8\mathrm{i} |\langle +_{B}|\psi\rangle|^{2}\Im \Big(\langle \Pi_{A}^{+}\rangle^{\psi,+_{B}}_{\mathrm{w}}\Big),
\end{eqnarray}
where $\langle \Pi_{A}^{+}\rangle^{\psi,+_{B}}_{\rm w}$ is the weak value of the projector $\Pi_{A}^{+}$ for a post-selected state $|+_{B}\rangle$ and pre-selected state $|\psi\rangle$ and $|\langle +_{B}|\psi\rangle|^{2}$ is the probability of successful post-selection. 

If $\hat{A}=\hat{\sigma}_{z}$ and $\hat{B}=\hat{\sigma}_{x}$, the commutation relation is  
\begin{eqnarray}\label{eq:com}
\langle\psi\vert(\hat{\sigma}_z\hat{\sigma}_x-\hat{\sigma}_x\hat{\sigma}_z)\vert\psi\rangle=2{\rm i}\langle\psi\vert\hat{\sigma}_y\vert\psi\rangle.
\end{eqnarray}
 By writing $\hat{\sigma}_x=2\Pi_{x}^{+}-\mathbb{1}$, $\hat{\sigma}_y=2\Pi_{y}^{+}-\mathbb{1}$ and  $\hat{\sigma}_z=2\Pi_{z}^{+}-\mathbb{1}$ where $\Pi_{x}^{+}=\vert +_{x}\rangle\langle +_{x}\vert$, $\Pi_{y}^{+}=\vert +_{y}\rangle\langle+_{y}\vert$ and $\Pi_{z}^{+}=\vert +_{z}\rangle\langle +_{z}\vert$ are the projectors with positive eigenvalue corresponding to the observables $\hat{\sigma}_{x}$, $\hat{\sigma}_{y}$ and $\hat{\sigma}_{z}$, respectively, from Eq.(\ref{eq:com}) it follows

\begin{figure*}[!t]
	\includegraphics[width=0.80\textwidth]{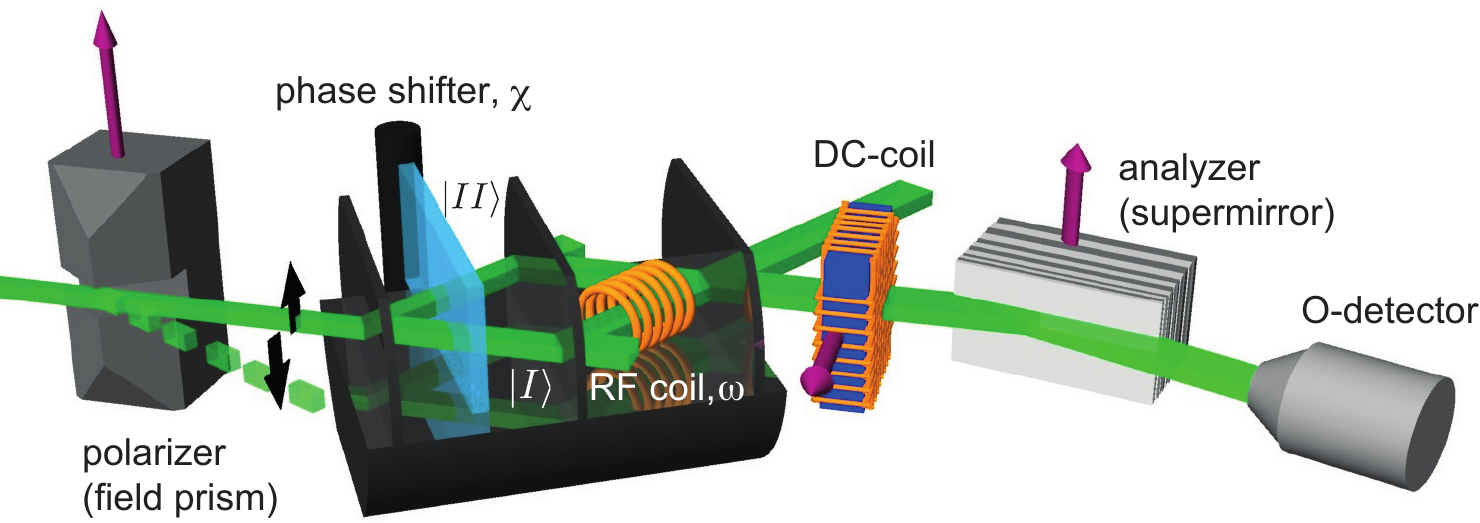}
	\caption{Illustration of the experimental setup for weak measurement of the path state. Polarized monochromatic neutrons enter the interferometer and are split into path $\ket{I}$ and $\ket{II}$ at the first interferometer plate. Preselection is achieved by adjusting the phase shifter plate and selecting a corresponding relative phase $\chi$ between the two paths. In path $I$, the neutron beam passes through a resonance frequency (RF) 
	spin-flipper, operating at frequency $\omega = 60$~kHz,  and is marked by a tiny energy-kick of $\hbar \omega$. After post-selection of the path state via projection in forward direction at the third interferometer plate, the spin direction is analyzed through the combined action of a direct-current spin-flipper (DC) and the supermirror (SM) in the O beam. The neutron counts are registered in a He-3 counter tube. Further details of setup and analysis of the signals are placed in Methods Section.}
	\label{fig:weak_path}
\end{figure*}

\begin{eqnarray}
\label{finalcom}
4\vert\langle+_{x}\vert\psi\rangle\vert^2\,\Im\Big(\langle\Pi^+_z\rangle^{\psi,+_{x}}_{\rm w}\Big)=-2\lvert\braket{+_{y}}{\psi}\rvert^2 + 1,
\label{eq:comm_rel_WV}
\end{eqnarray}
where $	\langle\Pi^+_z\rangle^{\psi,{+_{x}}}_{\rm w}={\langle{+_{x}}\vert\Pi^+_z\vert \psi\rangle}/{\langle{+_{x}}\vert\psi\rangle}$ is the weak value of $\Pi^{+}_{z}$ given the pre- and post-selected states $|\psi\rangle$ and $|{+_{x}}\rangle$, respectively. Thus, the imaginary part of a single weak value $\langle\Pi^+_z\rangle^{\psi,{+_{x}}}_{{\rm w}}$ is amenable to test the left-hand-side of the commutation relation in Eq.\,(\ref{eq:com}). For determining the quantity  on the right-hand side one requires to separately measure the post-selected probability $\lvert\braket{+_{y}}{\psi}\rvert^2$. 

\section*{Experiment}

To determine the weak value, an established neutron interferometric setup \cite{Rauch74,Sponar15Weak,Denkmayer17,Geppert18}, depicted in Fig.\,\ref{fig:weak_path}, is applied. The system is prepared (pre-selected)
and post-selected in the \emph{path} states: 
\begin{equation}
	\begin{split}
		&\ket{\psi_{\rm i}(\chi)} = \frac{1}{\sqrt{2}}(\ket{I}+e^{-{\rm i}\chi}\ket{II}) \\
		&\ket{\psi_{\rm f}} \equiv |+_{x}\rangle = \frac{1}{\sqrt{2}}(\ket{I}+\ket{II}),
	\end{split}
\end{equation}
where $\ket{I}$ and $\ket{II}$ denotes the path eigenstates of the Pauli path observable given by $\sigma_{z}=|I\rangle\langle I| -|II\rangle\langle II|$.
The imaginary part of the weak value of the path projector $\Pi_{1}$, is expressed as $\Im\left(\langle \Pi_1 \rangle^{\psi_{\rm i},+_{x}}_{\rm w}\right)= \sin \chi/(2+2\cos \chi)$ and probability of successful post-selection $\lvert\braket{+_{x}}{\psi_{\rm i}(\chi)}\rvert^2 =4\cos^{2}(\chi/2)$, with $|+_{x}\rangle = \left(\ket{I}+\ket{II}\right)/\sqrt{2}$. 
The quantity on the right-hand side of Eq.\,(\ref{finalcom}) is determined by the post-selected probability as $-2\lvert\braket{+_{y}}{\psi_{\rm i}(\chi)}\rvert^2 + 1= \sin \chi$, with $|+_{y}\rangle = \left(\ket{I}+{\rm i}\ket{II}\right)/\sqrt{2}$.

\subsection*{Experimental setup}

The experiment was performed at the S18 interferometer beam line at the research reactor of the Institute Laue-Langevin (ILL), in Grenoble, France. The setup is depicted in Fig.\,\ref{fig:weak_path}. A monochromatic neutron beam with a wavelength of $\lambda=1.92\,\angstrom$ passes through a magnetic prism, which produces two divergent, polarised beams with \emph{spin} states $\ket{\uparrow_{ z}}$ and $\ket{\downarrow_{ z}}$, respectively. For our purpose, only the beam with spin state $\ket{\uparrow_{ z}}$   is adjusted to fulfill the Bragg condition of the down-stream silicon perfect crystal interferometer of triple Laue geometry. The other beam is transmitted in forward direction and absorbed by a slab of cadmium to reduce the background noise in the detectors. The first crystal plate splits the beam into two paths $|I\rangle$ and $|II\rangle$. After passing through the first beam-splitter, a phase shifter is inserted into the interferometer to tune the relative phase $\chi$  between path $I$ and path $II$ and prepare the pre-selected path state $\ket{\psi_{\rm i}(\chi)}$. A resonance frequency (RF) spin manipulator is placed along path $|I\rangle$ to implement the weak measurement by small manipulation of the spin degree of freedom. 
Both beams are recombined at the third interferometer plate.


%
Of the two outgoing beams of the interferometer, only the neutrons in the $O$ beam are selected for analysis corresponding to the post-selected path state $\ket{\psi_{ f}}= |+_{x}\rangle$. The transmitted O beam passes through a direct current (DC) spin rotator and a polarizing CoTi multilayer array (henceforth referred to as supermirror) to perform spin analysis in the O beam. The refracted H beam is used as a reference. The intensities in both outgoing beams are measured using $^3$He counting tubes with a high detection efficiency (over \,97\,\%). 
The detector for the O beam has a diameter of \SI{6}{\mm} and is placed directly at the outgoing window of the supermirror. For the used wavelength a time resolution of $\SI{3}{\us}$ is achieved.


\subsection*{Weak value extraction}

Given the pre- and post-selected path states $\ket{\psi_{\rm i}(\chi)}$ and $\ket{+_{x}}$, respectively, we performed the measurement of the weak value $\langle \Pi_1 \rangle^{\psi_{\rm i},+_{x}}_{\rm w}  $ of the path projector $\Pi_{1}=|I\rangle\langle I|$. 
The weak interaction is implemented by use of the RF spin manipulator in path $|I\rangle$. Its effect can be represented by a unitary evolution as $\mathcal{U}_{\rm{int}}=\Pi_1\otimes \,\mathcal{U}_{\rm{RF}}(t,\alpha,\omega,\delta) + \Pi_2$. 
Here, $\alpha$ is the spin-rotation angle, $\omega$ the frequency of the electromagnetic RF field and $\delta$ an arbitrary phase of this RF field. The parameter $\alpha$ is directly dependent on the magnetic field strength and a sufficiently small value of it warrants the required weak measurement criteria. 
In this experiment, $\alpha=\pi/9$ and $\omega=60$ kHz were taken. As explained in detail in 
Methods, the weak value $\langle\Pi_{1}\rangle^{\psi_{\rm i},+_{x}}_{\rm w}$ can be extracted from the time-dependent intensity at the O detector that is given by
\begin{equation}
	I(t) = \frac{1}{2}\lvert \braket{+_{x}}{\psi} \rvert^2 \Big( 1 + \alpha \Im( e^{{\rm i} \omega t } \langle \Pi_1 \rangle^{\psi_{\rm i},+_{x}}_{\rm w} )  \Big),
   \label{eq:time_dep_intensitymain}
\end{equation} 

\begin{figure}[!b]
	\centering
	\includegraphics[width=0.44\textwidth]{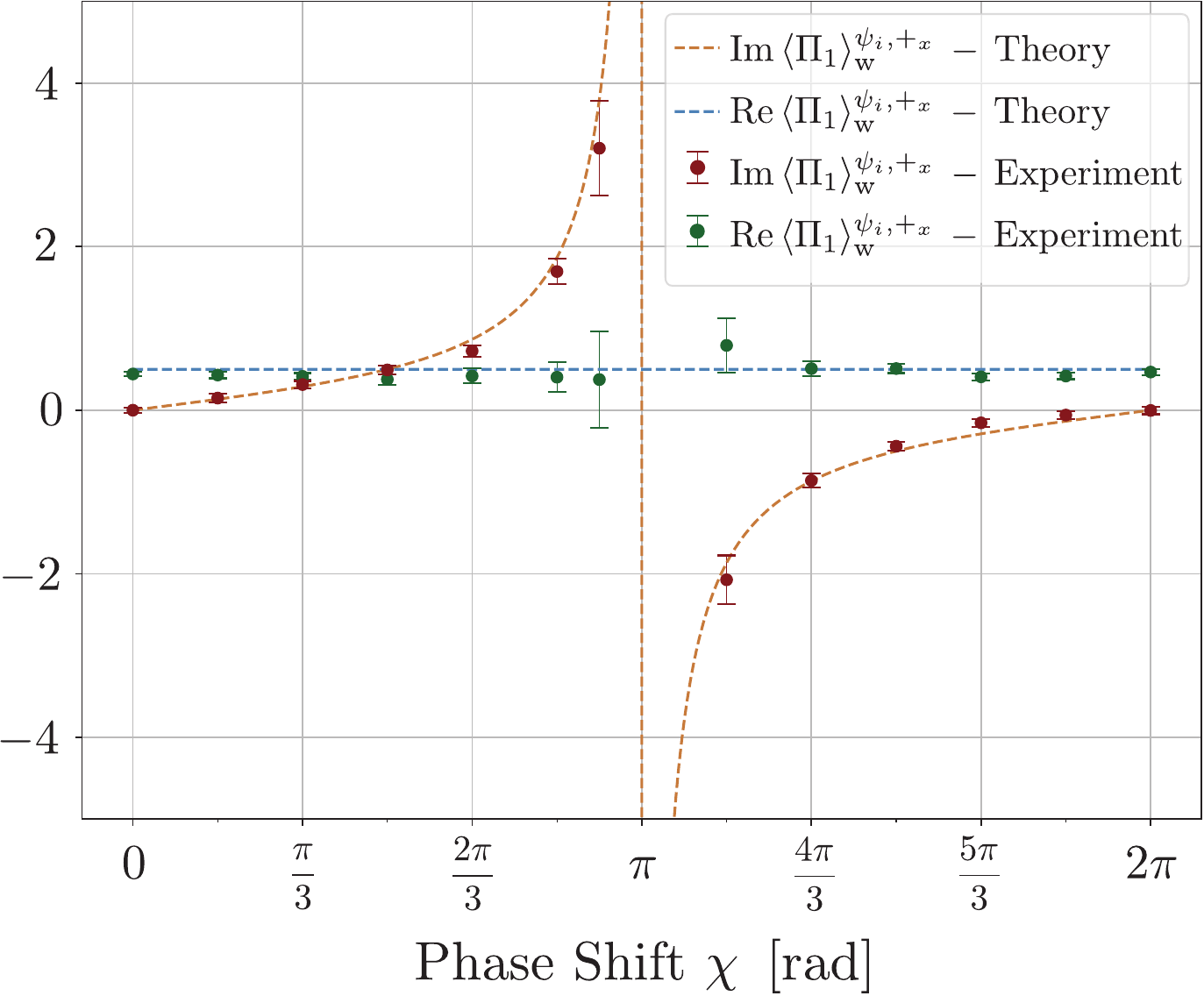}
	\caption[]{
		Experimental results of the real and imaginary parts of weak value  $\langle \Pi_1 \rangle^{\psi_{i},+_{x}}_{\rm w}  $ are plotted versus $\chi$. The solid dots denote the least-square fits to the data and the error bars represents unit standard deviation.} 
	\label{fig:expResults_WV}
\end{figure}
%
%
By switching off the RF field, two post-selected probabilities $\lvert \braket{+_{x}}{\psi} \rvert^2 $ and $\lvert\braket{+_{y}}{\psi_{\rm i}(\chi)}\rvert^2$ are measured. The former resembles an empty interferogram while the latter  
corresponds to an empty but by $\chi_{\rm{add}} = -\pi/2$ shifted interferogram \cite{RauchBook}. 

The weak value is in general a complex number and can be expressed in polar form as $\langle \Pi_1 \rangle^{\psi,+_{x}}_{\rm w} = A e^{\mathrm{i}\varphi}$. The amplitude $A$ and phase $\varphi$ are retrieved from a sinusoidal fit to the time-dependent intensity. The reference phase $\varphi_{\rm ref} = 0 $ is obtained for the particular case when pre- and post-selected states coincide, i.e., $\chi=0$. In Fig.\,\ref{fig:expResults_WV}, we present the detailed analysis of the results of the weak measurement of $\langle \Pi_1 \rangle^{\psi,+_{x}}_{\rm w}$. For comparison, the experimental values of the real and the imaginary part of $\langle \Pi_1 \rangle^{\psi,+_{x}}_{\rm w}$ are separately plotted along with the theoretical predictions $\langle \Pi_1 \rangle^{\psi,+_{x}}_{\rm w}=1-\frac{1}{1+e^{{\rm i}\chi}}$. As can be seen from the experimental results, the theoretical predictions are reproduced evidently.

 \begin{figure}[!t]
	\centering
	\includegraphics[width=0.45\textwidth]{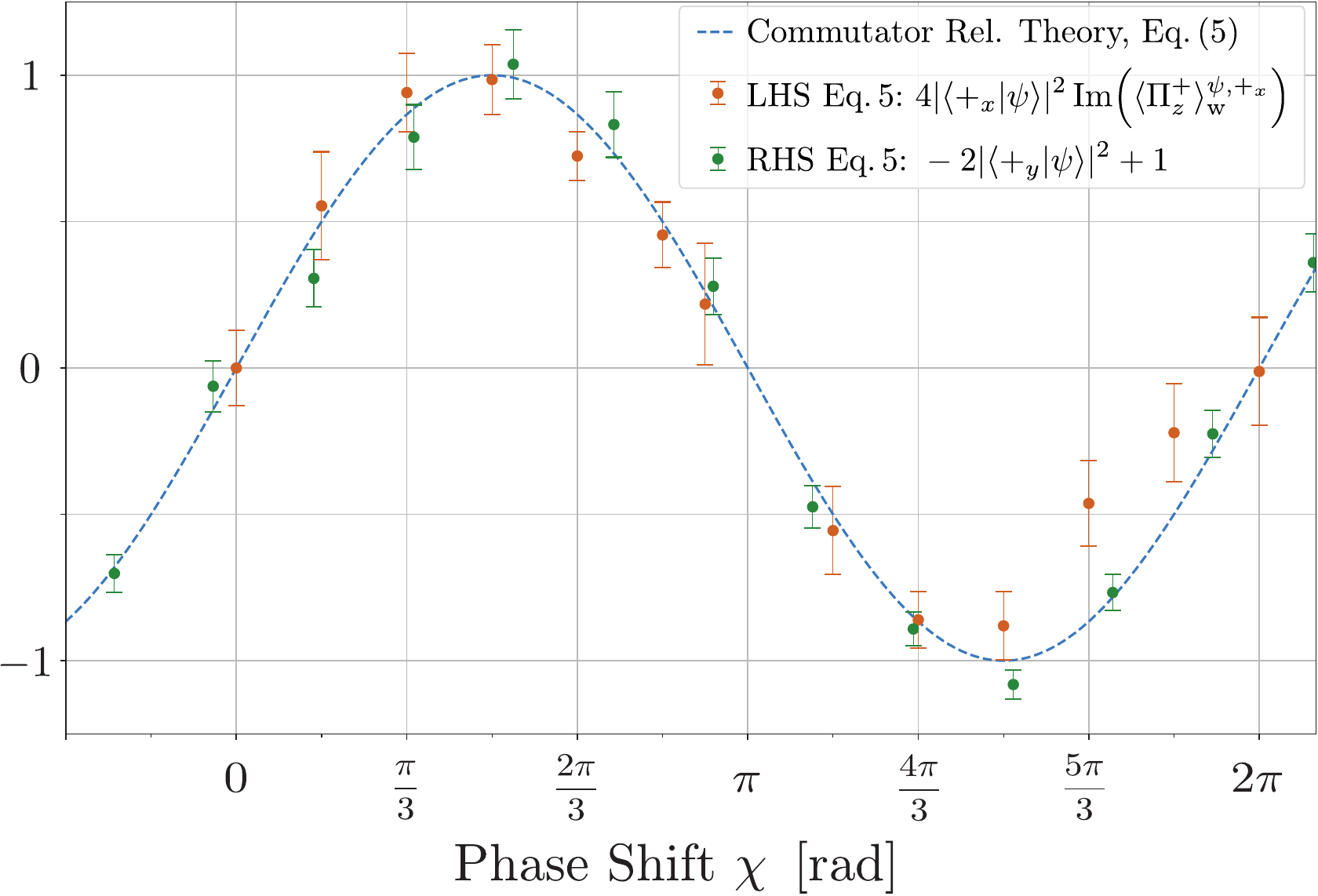}
	\caption[]{
		Experimental results of left-hand side (orange) and right-had side (green) of the commutation relation in Eq.\,({\ref{eq:comm_rel_WV}}) are plotted as a function of phase-shift $\chi$. The results are in excellent agreement with the relevant theoretical prediction (dotted blue). The error bars represent one standard deviation.  }
	\label{fig:expResults_commutatorRel}
\end{figure}



Finally, the results of all three measurements, namely
$\braket{+_{x}}{\psi} \rvert^2 $, $\lvert\braket{+_{y}}{\psi_{\rm i}(\chi)}\rvert^2$ and 
$\operatorname{Im}(\langle\Pi_{1}\rangle^{\psi_{\rm i},+_{x}}_{\rm w})$, 
are combined in Fig.\,\ref{fig:expResults_commutatorRel}, for the direct test of the commutation relation.
The left- and right-hand side related measurements of the commutation relation (Eq.\,(\ref{eq:comm_rel_WV})) are shown by orange and green data points respectively.
The result is in good agreement with the relevant theoretical prediction (dotted blue), verifying the canonical commutation relation, as expressed in Eq.\,(\ref{eq:comm_rel_WV}).
%

\section *{Summary and Discussion}

The results above confirm the canonical commutation relation by weak measurement of the path-qubit observable in a neutron interferometer. 
Accordingly, from the quantum foundational perspective, our experiment thus provides a genuine direct test of one of the fundamental tenets of quantum theory. This test is as fundamental as the direct measurement of a quantum wave function by Lundeen \emph{et al.} \cite{lundeen11}. Heisenberg's uncertainty relation may be regarded as a direct consequence of the canonical commutation relation and  several formulations of it have been tested experimentally \cite{Shull,Yuen,Breitenbach,erhart12,steinberg12}. 

At this point we want to emphasize, that our presented scheme is fundamentally different compared to prior experiments studying the commutation relation  \cite{Hasegawa1997, Wagh1997, Kim2010}. In these experiments the non-commutativity of Pauli spin matrices in terms of different sequences of rotations (on the same initial state) was utilized. However, a genuine test of the canonical commutation relation should unambiguously determine the expectation value of the non-Hermitian product of two non-commuting observables as it occurs in the commutation relation. As presented here, the direct test of the commutation relation boils down to the experimental determination of the imaginary part of a single weak value of a suitable path observable. The weak value has provided a multitude of practical and conceptual implications \cite{duck,av91,resch,pusey,dressel,dressel12,nirala2019,vaid13,ah13,pigeon1, ritchie,pryde,hosten,jwvexp,starling,steinberg11,lundeen11,danan,piacentini,denk, xu,david}. The imaginary part of the weak value is more counter-intuitive than its real counterpart. The latter can be operationally interpreted as an idealized conditioned average value of the concerned observable in the zero measurement disturbance limit. On the other hand, the imaginary part is associated with back-action on the system due to the measurement itself \cite{dressel12}. This explanation conceptually fits well with our test of the commutation relation through the imaginary part of the weak value. Finally, it would be interesting to test the traditional position-momentum uncertainty relation following the approach developed here. 
The experimental method presented here could be expanded to more path markers in the interferometer and several path weak values could be measured simultaneously. Applying this procedure to a four-plate interferometer would even offer more than two path markers \cite{Geppert18}.

\begin{acknowledgements}
This work was financed by the Austrian Science Fund (FWF), Project No. P30677.
\end{acknowledgements}
\section*{Methods}
\subsection*{Detailed analysis of the intensity for the weak value extraction}

Given the pre-selected \emph{path} state denoted explicitly as $\ket{\psi_{\rm i}(\chi)} = \frac{1}{\sqrt{2}}(\ket{I}+e^{-\rm i\chi}\ket{II})$ and post-selected path state $\ket{\psi_{\rm f}} = \frac{1}{\sqrt{2}}(\ket{I}+ \ket{II})$, to determine the weak value of path observable $\Pi_1$, an interaction of the resonance frequency (RF) spin manipulator in path $|I\rangle$ is employed which can be represented by the unitary dynamics, denoted as  $\mathcal{U}_{\rm{int}}=\Pi_1\otimes \,\mathcal{U}_{\rm{RF}}(t,\alpha,\omega,\delta) + \Pi_2$.  The spin manipulation leads to an energy-shift by the amount $\Delta E = \pm\hbar \omega$ in the spin-flipped parts and can be written as a unitary operator:  
\begin{equation}
\mathcal{U}_{\rm{RF}}(t,\alpha,\omega,\delta)=\begin{pmatrix}
\cos(\frac{\alpha}{2}) & {\rm i} \sin(\frac{\alpha}{2}) e^{+{\rm i}(\omega t + \delta)} \\ 
{\rm i} \sin(\frac{\alpha}{2}) e^{-{\rm i}(\omega t + \delta)} & \cos(\frac{\alpha}{2})
\label{unitary_op_RF}
\end{pmatrix},
\end{equation}
where $\alpha$, $\omega$ and $\delta$ are the spin-rotation angle,  the frequency of the electromagnetic RF field and an arbitrary phase of this RF field. The energy-shifted contributions appear due to the off-diagonal terms. The parameter $\alpha$ is related to magnetic field strength and for a small value of $\alpha$ \eqref{unitary_op_RF} becomes:
\begin{equation}
\lim\limits_{\alpha \rightarrow 0} \,\mathcal{U}_{\rm{RF}} = \mathds{1} - {\rm i}\frac{\alpha}{2} \begin{pmatrix}
0 & e^{+{\rm i}(\omega t + \delta)} \\ 
e^{-{\rm i}(\omega t + \delta)} & 0
\label{unitary_op_RF_approx}
\end{pmatrix}.
\end{equation}
The total path-spin state of the in $\uparrow_z$-direction polarized neutrons after the first plate of the interferometer is 
\begin{equation}
\ket{\Psi}=\ket{\psi_{\rm i}(\chi)}\otimes\ket{\uparrow_{z}}.
\end{equation}
Spin related states and projectors are indicated by the $\uparrow$ symbol. 
After the interaction with the spin-modulator in path $I$, the state evolves to
$\ket{\Psi'} = \,\mathcal{U}_{\rm{int}}\ket{\Psi}$ and is subsequently projected onto the post-selected path state $\ket{\psi_{\rm f}}$ at the third crystal plate 
\begin{equation}
\ket{\Psi'} \rightarrow \ket{\Psi''} = \Pi_{\rm f}\,\mathcal{U}_{\rm{int}}\ket{\Psi}.
\end{equation}
where $\Pi_{\rm f} = \ket{\psi_{\rm f}}\bra{\psi_{\rm f}}$. Now, for our purpose, the spin analysis is performed for $\ket{\uparrow_x}$ which is implemented  by the action of the DC coil and the supermirror (SM). The joint (un-normalized) path-spin state is projected onto 
$P_{\uparrow_x} = \ket{\uparrow_x}\bra{\uparrow_x}$: 
\begin{equation}
 \ket{\Psi'''} = P_{\uparrow_x}\Pi_{\rm f} \,\mathcal{U}_{\rm{int}}\ket{\Psi}.
\end{equation}
\begin{figure}[!b]
	\centering
	\includegraphics[width=0.45\textwidth]{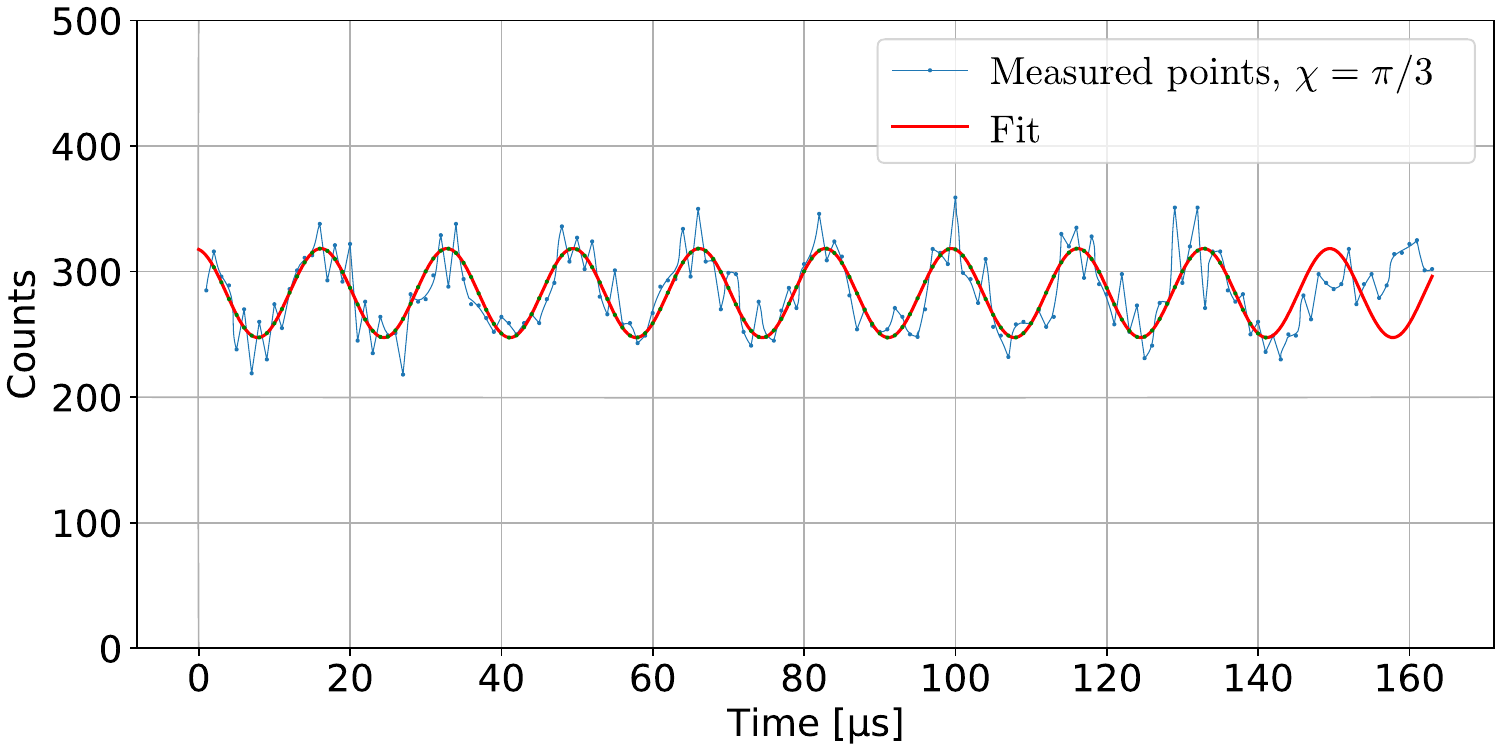}
	\caption[]{A typical time resolved intensity signal of the measurement campaign for the phase-shift $\chi=\pi/3$.}
	\label{fig:expResults_TOFRaw}
\end{figure}
Using \eqref{unitary_op_RF_approx} we get the approximation in the limit for small $\alpha$ (neglecting contributions of order $\mathcal{O}(\alpha^2)$ or higher):
\begin{equation}
\begin{split}
\ket{\Psi'''}  \approx\frac{1}{\sqrt{2}}\ket{\psi_{\rm f}}\otimes\ket{\uparrow_x}\,\bra{\psi_{\rm f}}\mathds{1} - {\rm i}\frac{\alpha}{2} (\Pi_1 e^{{\rm i}(\omega t + \delta)} )\ket{\psi_{\rm i}(\chi)}.
\end{split}
\end{equation}
Further rearrangement provides 
\begin{eqnarray}
\ket{\Psi'''}&=&\frac{1}{\sqrt{2}}(\ket{\psi_{\rm f}}\otimes\ket{\uparrow_x})\,\braket{\psi_{\rm f}}{\psi_{\rm i}(\chi)}\nonumber\\&&\left( 1 - \, {\rm i}\frac{\alpha}{2} (\langle\Pi_1\rangle_{w}^{\psi_{\rm i},\psi_{\rm f}} e^{{\rm {\rm i}}(\omega t + \delta)}  ) \right).
\end{eqnarray}
The time-dependent intensity $I_{\rm{ideal}}(t)=|\Psi'''|^{2}$ in the ideal scenario can then be written as 
\begin{equation}
I_{\rm{ideal}}(t)=\frac{1}{2}\lvert\braket{\psi_{\rm f}}{\psi_{\rm i}(\chi)}\rvert^2\left( 1+\alpha \operatorname{Im}(\langle\Pi_1\rangle_{w}^{\psi_{i},\psi_{\rm f}} e^{{\rm i}(\omega t + \delta)})\right),
 \label{I_t_ideal}
\end{equation}
which is dependent on $\chi$. With $\ket{\psi_{\rm f}}\equiv \ket{+_{x}}$, $\langle \Pi_1 \rangle^{\psi_{\rm i},+_{x}}_{\rm w}= (1+e^{+{-\rm i}\chi})^{-1}$ can be extracted from the intensity analysis.    
By switching off the RF spin modulator $(\alpha=0)$, 
the signal for post-selected probabilities $\lvert\braket{+_{x}}{\psi_{\rm i}(\chi)}\rvert^2 =(1+\cos \chi)/2$ and $\lvert\braket{+_{y}}{\psi_{\rm i}(\chi)}\rvert^2 =(1-\sin \chi)/2$ in Eq.\,(5) can be measured. For the latter, in practice this is done by applying an additional 
phase shift of $\chi_{\rm{add}}= -\pi/2$ to the readings from the measurement of the empty interferogram.

\subsection*{Data analysis} 
In experiments, there inevitably occurs a loss of coherence in the signal. Therefore, the intensity has to be corrected by carefully taking into account the incoherent contributions. The amount of this incoherence can be accounted for by the contrast or fringe visibility parameter $\eta \in [0,1]$. The intensity then reads as
\begin{equation}
I_{\rm{real}}= \eta \, I_{\rm{ideal}} + (1-\eta) I_{\rm{inc}}
\label{adaptC},
\end{equation}
where $I_{\rm{inc}}$ stands for its incoherent part.
By considering an empty interferometer, we quantify $I_{\rm{ideal}}^{\rm{empty}}$ as
\begin{eqnarray}
I_{\rm{ideal}}^{\rm{empty}} &=& \lvert\braket{\psi_{\rm f} }{ \psi_{\rm i}}\rvert^2 = \lvert \underbrace{\braket{I}{I}}_{\Psi_1} + \underbrace{e^{{\rm i}\chi}\braket{II}{II}}_{\Psi_2} \rvert^2\nonumber\\&=& \, \lvert\Psi_1\rvert^2 + \lvert\Psi_2\rvert^2 + 2 \operatorname{Re}(\Psi_1^\ast \Psi_2).
\label{ideal_empty}
\end{eqnarray}
\begin{figure}[!b]
	\centering
	\includegraphics[width=.45\textwidth]{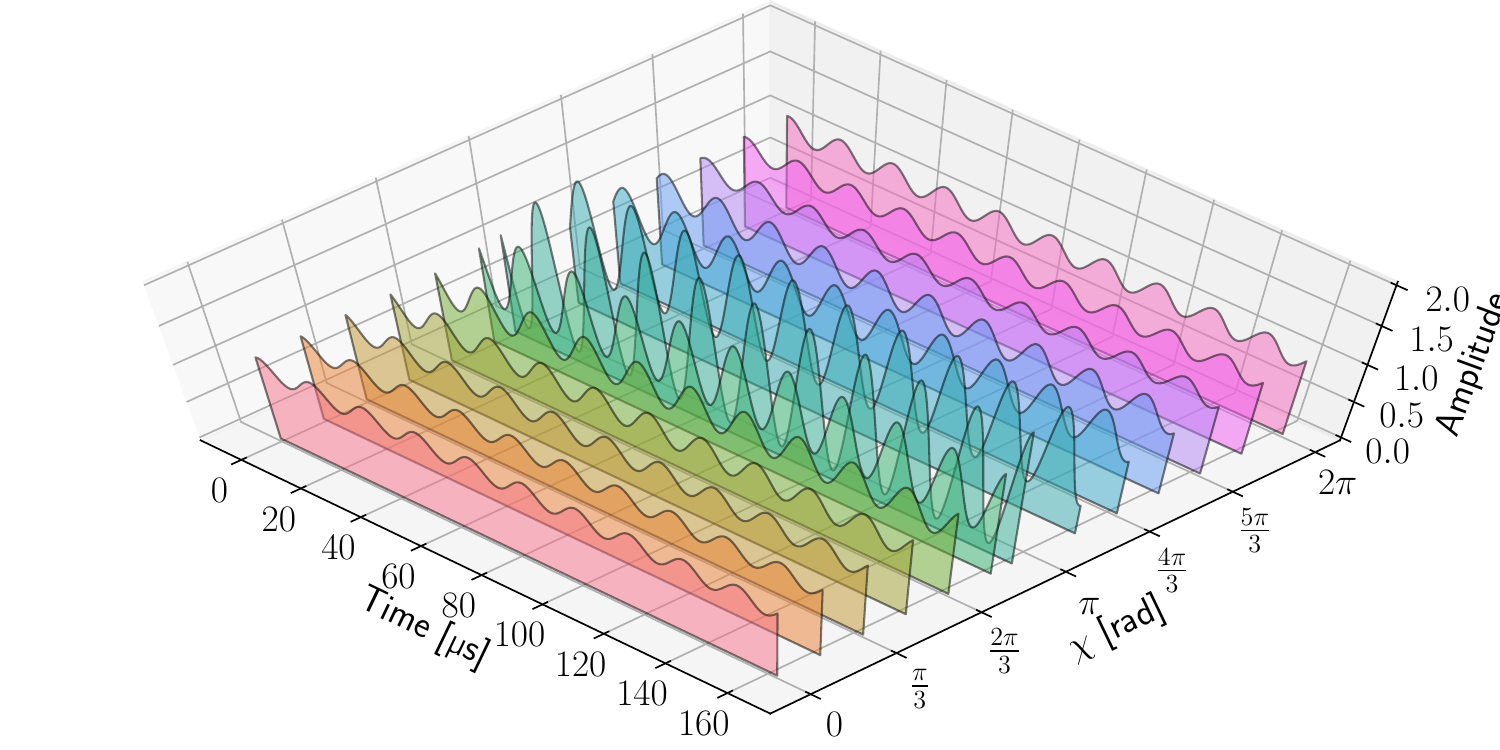}
	\caption[]{Collected time resolved  (normalized) intensity signals for varying phase shifter position $\chi$. }
	\label{fig:expResultsTOFSignalsCollected}
\end{figure}

In an ideal scenario we expect full coherence of the interference term $2\operatorname{Re}(\Psi_1^\ast \Psi_2)$, i.e., $\eta=1$. But, in real experiment, partial coherence can be obtained, so that in our experiment $I_{\rm{real}}^{\rm{empty}}$ is quantified as
\begin{equation}
\begin{split}
I_{\rm{real}}^{\rm{empty}} \,\,&= \,\, \lvert\Psi_1\rvert^2 + \lvert\Psi_2\rvert^2 + 2 \,\eta \operatorname{Re}(\Psi_1^\ast \Psi_2)\\&= \eta\,\lvert \Psi_1 + \Psi_2\rvert^2 + (1-\eta)\left(\lvert\Psi_1\rvert^2 + \lvert\Psi_2\rvert^2\right). \\ 
\end{split} \label{eq:I_C_2path}
\end{equation}

\begin{figure}[!t]
	\centering
	\includegraphics[width=0.4\textwidth]{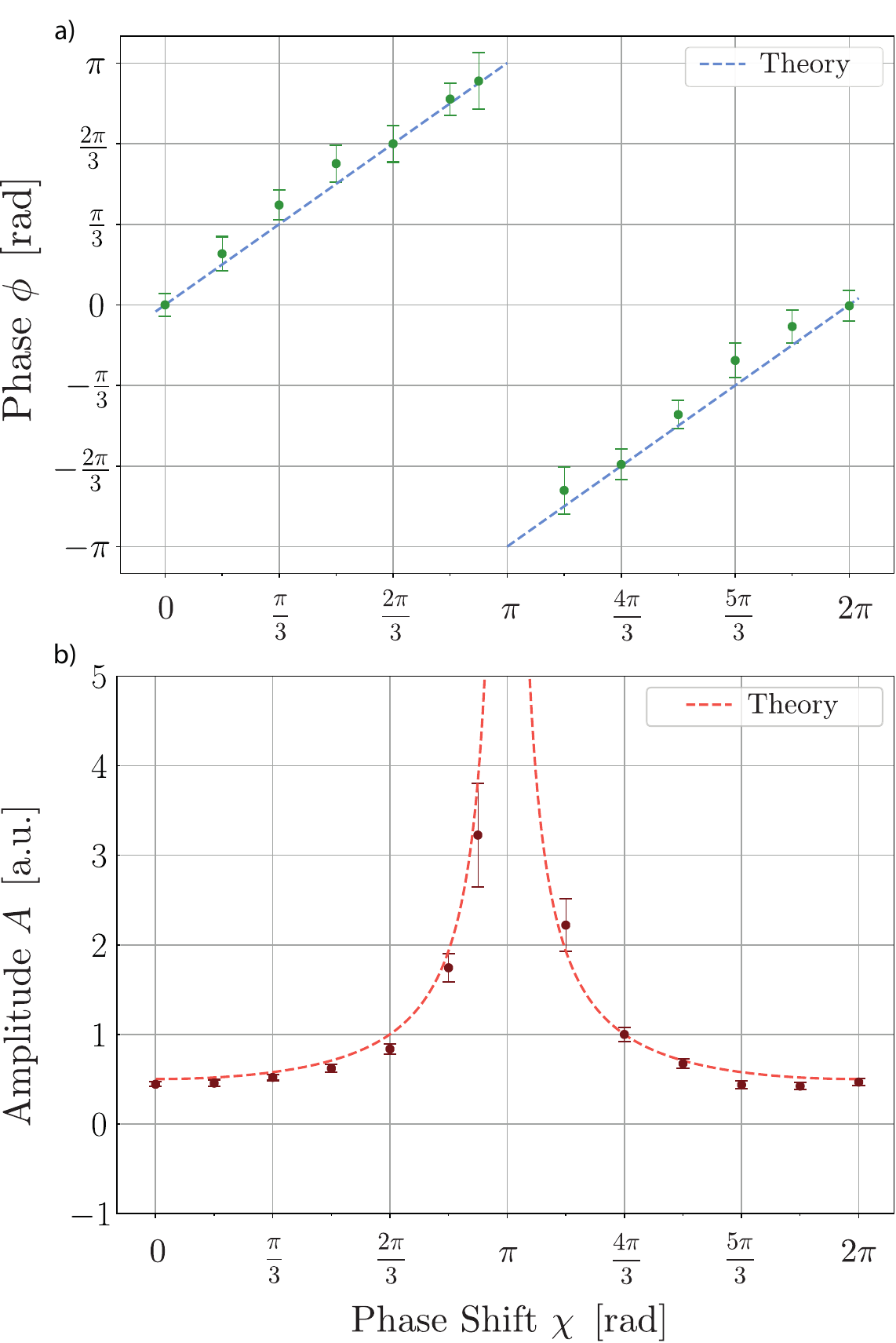}
	\caption[]{a) Phase and b) Amplitudes of the weak value measurement extracted from the time resolved intensity signals.}
	\label{fig:expResults}
\end{figure}
Identifying the incoherent part as $I_{\rm{inc}} = \lvert\Psi_1\rvert^2 + \lvert\Psi_2\rvert^2$, it is straightforward to see that Eq.\,\eqref{eq:I_C_2path} is equivalent to Eq.\,\eqref{adaptC}. The time-dependent intensity in our experiment  can then be written as
\begin{equation}
	\begin{split}
		I_{\rm{real}}(t) &= \frac{\eta}{2}\lvert\braket{\psi_{\rm f}}{\psi_{\rm i}(\chi)}\rvert^2\left( 1+\alpha \operatorname{Im}(\langle\Pi_1\rangle_{\rm w} e^{{\rm i}(\omega t + \delta)})\right)\\
		&\quad + (1-\eta)\Bigg(\underbrace{\frac{1}{8}\left(1-\alpha\sin(\omega t + \delta)\right)}_{\lvert\Psi_1\rvert^2:=I_1(t)} + \underbrace{\frac{1}{8}}_{\lvert\Psi_2\rvert^2:=I_2(t)}  \Bigg),
	\end{split}
\end{equation}
%
$\lvert\Psi_1\rvert^2$ and $\lvert\Psi_2\rvert^2$ are intensities for the isolated path $I$ and $II$, respectively, which are measured separately to apply the correction. An example of such a signal is shown in Fig.\,\ref{fig:expResults_TOFRaw}. The visibility $\eta$ is extracted from an interferogram of an empty interferometer by fitting according to Eq.\,\eqref{eq:I_C_2path}. In our measurements the average value $\eta$ is $0.79$.
%

Now, the ideal intensity  is the one that is the measured intensity minus contributions from independent (decoherent) intensities

\begin{equation}
I_{\rm{ideal}}(t)= \frac{1}{\eta}\bigg(I_{\rm{real}}(t)-(1-\eta)\left(I_1(t) +  I_2(t)\right)\bigg).
\label{eq:I_corrected}
\end{equation}
Time-resolved signals have been recorded for several phase shifter settings in the range $0\leq\chi\leq 2\pi$. These have been corrected according to Eq.\,(\ref{eq:I_corrected}), fitted and normalized and are shown in Fig.\,\ref{fig:expResultsTOFSignalsCollected}.

The weak value for the path projector is  a complex number and can generally be  written in the polar form $\langle \Pi_1 \rangle_{\mathrm{w}} = A e^{{\rm i}\varphi}$. So it follows from Eq.\,\eqref{I_t_ideal} that, its amplitude and phase can be obtained from sine fits to the time-dependent intensities:
\begin{equation}
I_{\rm{ideal}}(t)=\frac{1}{2}\lvert\braket{\psi_{\rm f}}{\psi_{\rm i}(\chi)}\rvert^2\big( 1+\alpha A\sin(\varphi - \omega t - \delta)\big).
\label{Int_ideal_final}
\end{equation}
Here $\delta$ is an arbitrary additional constant phase.
The modulation in Eq.\,\eqref{Int_ideal_final} show a reduced 
fringe visibility, too. The retrieved values for the amplitudes $A$ have therefore to be corrected with $\eta$, according to $A_{\rm w} = A / \eta$ to obtain the amplitudes of the path projector weak values. 
%

%
At the phase shifter setting $\chi=0$, the weak value
$\langle \Pi_1 \rangle_{\rm w}= (1+e^{-{\rm i}\chi})^{-1}$
becomes real and exactly $\nicefrac{1}{2}$. So the phase angle $\varphi$ is equal to zero. 
This serves as a reference in the measurement to retrieve the phase angle $\varphi$ for the other $\chi$ settings.
A plot that collects the retrieved amplitudes $A_{\rm w}$ and phases $\varphi$ for the weak value of the path projector $\langle\Pi_1\rangle_{\rm w}$ is shown in Fig.\,\ref{fig:expResults}.

%
The errors in all plots are from propagation of the standard deviations of the fit errors for the time dependent intensity and reference interferogram signals and the systematic effects due to uncertainties in the adjustment of the rotation angle $\alpha$.

\end{document}